\documentclass[prb,aps,twocolumn,showpacs]{revtex4}

\usepackage[usenames]{color}
\usepackage{graphicx}%
\usepackage{dcolumn}
\usepackage{amsmath}

\makeatletter
\def\btt#1{\texttt{\@backslashchar#1}}%
\DeclareRobustCommand\bblash{\btt{\@backslashchar}}%
\makeatother

\begin{document}

\preprint{HEP/123-qed}

\title[Short Title]{Magnetism of Ru and Rh thin films on Ag(001) substrate}

\author{S. Meza-Aguilar, R. E. F\'elix-Medina, M. A. Leyva-Lucero}
\affiliation{%
Escuela de Ciencias Fisico-Matem\'aticas, Universidad Aut\'onoma de Sinaloa,
Bldv. de las Americas y Universitarios, Ciudad Universitaria, Culiac\'an Sinaloa,
CP 80010, M\'exico.
}%

\date{\today} 

\begin{abstract}
In a very recent x-ray magnetic circular dichroism experiment concerning with Ru and Rh impurities and metal films on Ag(001) substrate, no local magnetic moments were displayed in direct contradiction with previous theoretical works. It is thought that there can be three main reasons for this inconsistency: relaxation, alloying and many-body effects. Some of the above-mentioned systems are studied by using a first-principles method in which relaxation and alloying are taken into account, even so magnetism is still obtained. For low-coverage systems, high magnetic moments in both Ru ($\sim$ 2.49 $\mu_{B}$) and Rh ($\sim$ 2.00 $\mu_{B}$) are obtained. Naturally, as the coverage is increased the magnetic moments are approached to zero. Also, it is noticed that the relaxation distances are increased by magnetism, which in turn is decreased by alloying. The behavior of the magnetic properties is explained in terms of Stoner model.
\end{abstract}

\maketitle

\section{Introduction}

Low-dimensional systems involving transition metals (TM) is the subject of an extraordinary research activity which is driven by both fundamental and technological interest. The complexity of this field is illustrated by considering that frequently the experimental\cite{Honolka,Schmitz,Li,Mulhollan,Liu,Cox1,Cox2,Chang,Beckmann,Lin,Chado} and theoretical\cite{Zhu,Eriksson,Blugel92,Wu,Blugel95,Turek,Wildberger,Stepanyuk1,Stepanyuk2,Bazhanov,Cabria,Bellini} works are in contradiction each other. For example, the theoretical possibility of ferromagnetism in 4$d$ TM thin films on noble substrates was originally reported by Zhu\cite{Zhu} and Eriksson\cite{Eriksson}, then several experimental works were performed\cite{Li,Mulhollan,Liu}: some of those with the intention of creating\cite{Li} that kind of systems in the laboratory and others with the idea of searching for magnetism\cite{Mulhollan,Liu}. However, magnetism was not found experimentally. This lack of accord was regarded as a consequence of that neither relaxation nor pseudomorphism were taken into account in the theoretical models. In this sense, Wu and Freeman\cite{Wu}, using a full potential method which took into consideration relaxation by means of total energy minimization, presented a comparison between magnetic and paramagnetic states as a function of the Ru(Rh)-Ag interlayer distance. In the same direction Bl\"ugel\cite{Blugel95} ignoring the force minimization reported the dependence of magnetism as a function of the layer coverage. Afterward, Turek\cite{Turek} {\it et al.} using an {\it ab initio} method and making allowance for mixing with the Ag atoms, obtained the magnetic moment as a function of the coverage and the mixing with the Ag substrate. Besides, experimental works done by Chang\cite{Chang} {\it et al.} and Beckmann\cite{Beckmann} {\it et al.} presented the possibility of clusters formation in the surface, and due to this the vanishing of magnetism. Following these  experimental works Stepanyuk\cite{Stepanyuk1} {\it et al.} studied small and mixed $4d$ clusters on Ag(001) surface, and in according with other theoretical works, they found magnetic moments in Ru and Rh atoms. Much more recently, Honolka\cite{Honolka} {\it et al.}, using x-ray magnetic circular dichroism obtained no magnetic signal in Ru and Rh thin films on Ag(001) in contradiction with theoretical works\cite{Eriksson,Blugel92,Blugel95}. Also, Chado\cite{Chado} {\it et al.} using STM images reported absence of ferromagnetism in a study of Rh atoms deposited on Au(111) substrate.

The purpose of this paper is to give an advance in the comprehension of the subtle interplay between magnetism and geometry in Ru and Rh thin films on Ag(001). This specific problem is expected to be particularly interesting from a fundamental point of view due to the current inconsistency between theory and experiment\cite{Honolka}. In particular, it seems that two out of three mean reasons for this disagreement have been ruled out. 

The remainder of the paper is organized as follows. In the next section a brief outline of the {\it ab initio} method used for the calculations is given. The Results obtained for paramagnetic state and the magnetism in Ru and Rh thin films on Ag(001) fcc substrate are presented and discussed in section 3. There have been considered effects that in previous works have not been reckoned. Finally, section 4 summarizes the main conclusions.

\section{Theoretical method}

\begin{figure}[t!]
\includegraphics[scale=0.15]{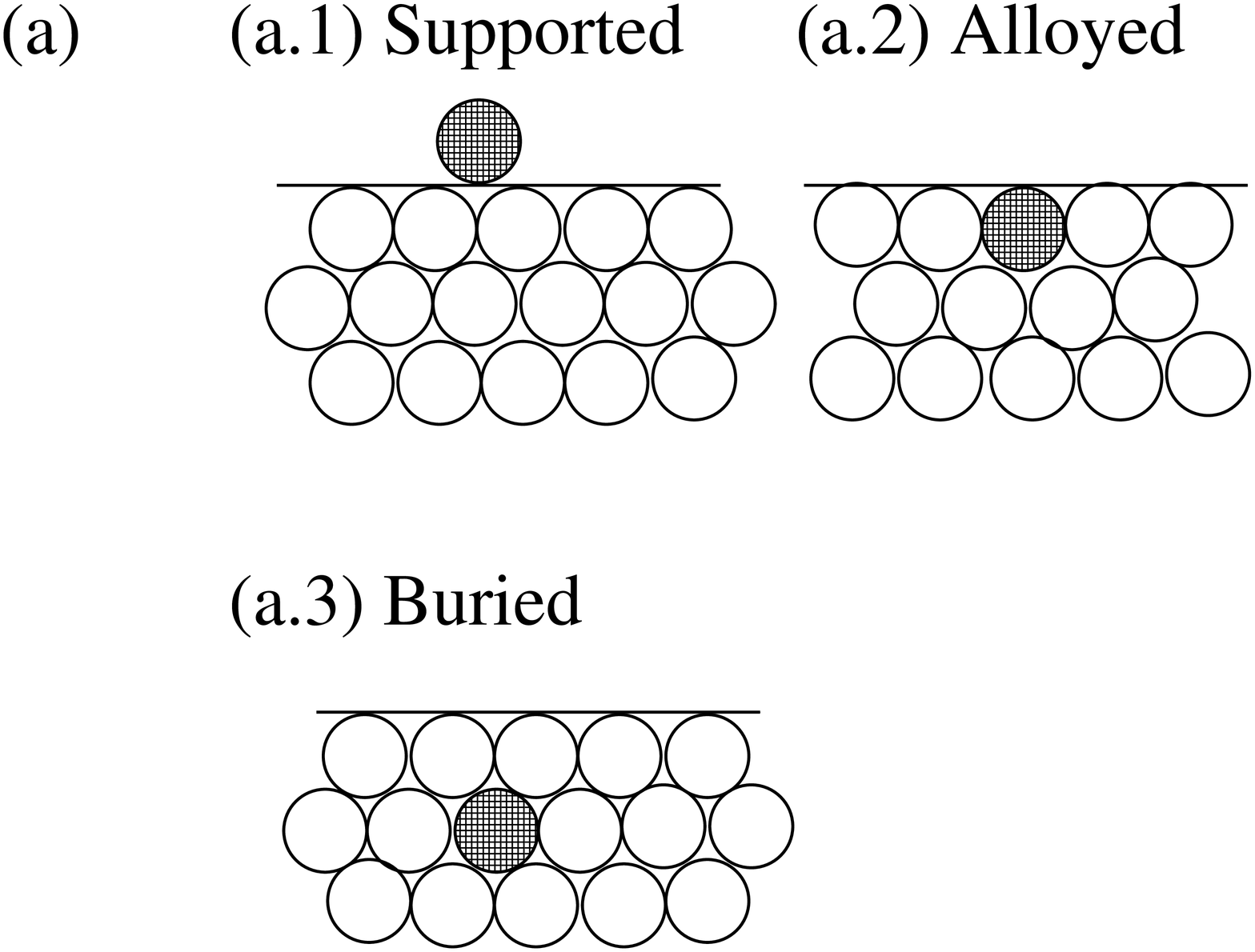}
\vspace{0.50cm}

\includegraphics[scale=0.15]{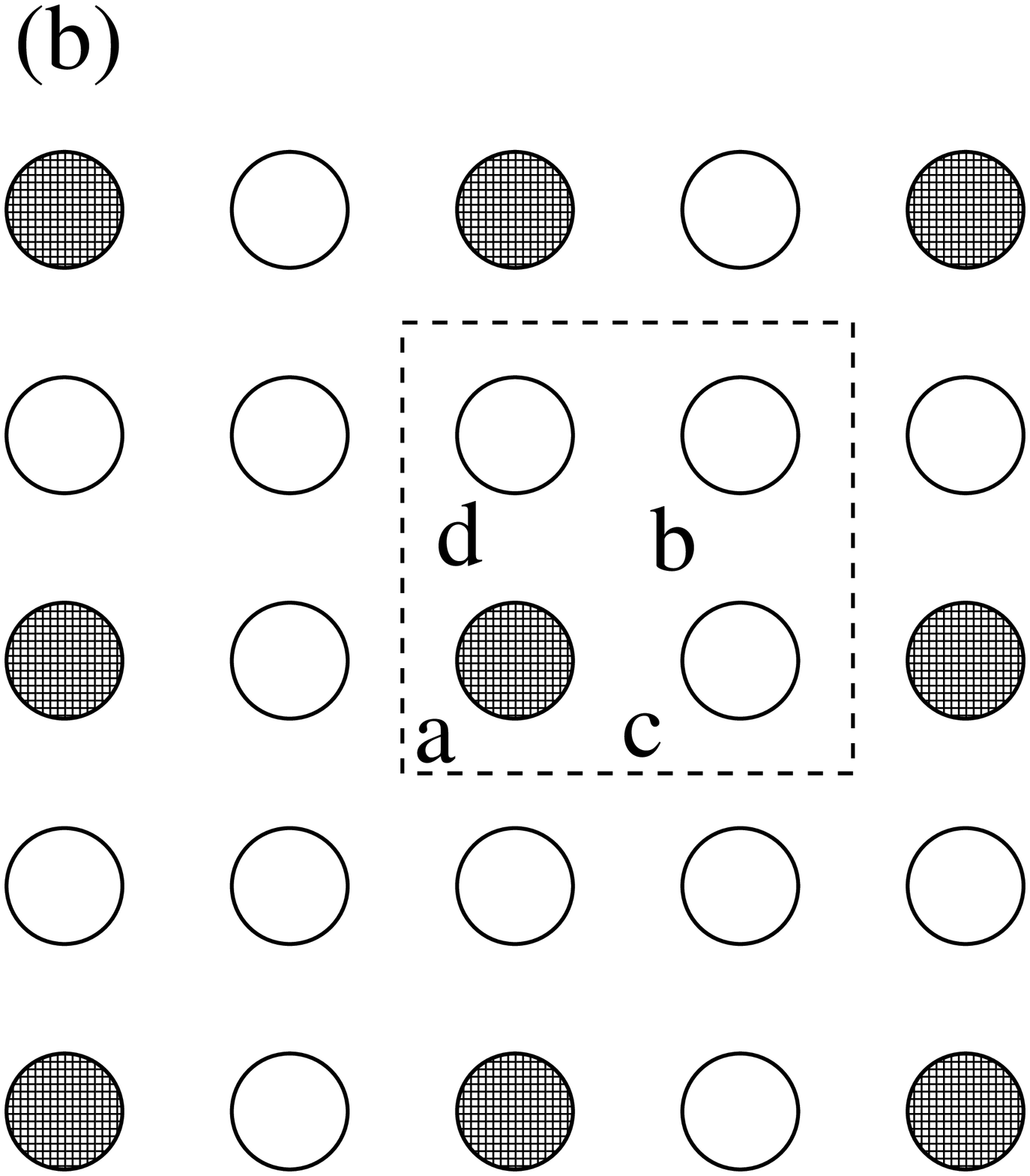}
\hspace{0.70cm}
\includegraphics[scale=0.15]{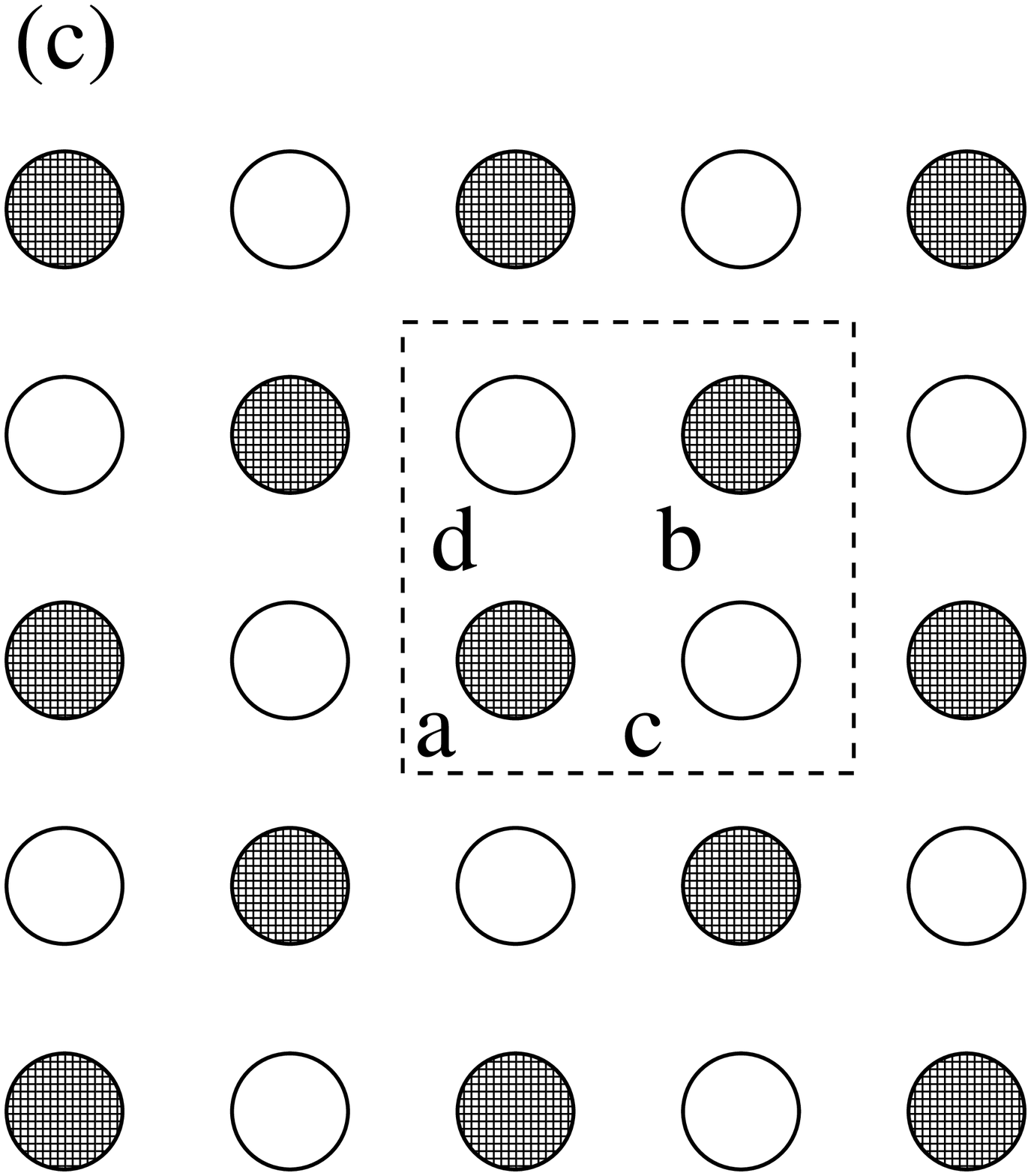}
\caption{\label{fig:Geometry}
Scheme of the geometrical configurations. (a) The systems are modelled in 3 cases, (a.1) supported, (a.2) alloyed and (a.3) buried atoms. Positions in the unit cell consist of 4 sites labelled by $a$, $b$, $c$ and $d$ inside the dashed line, see (b) and (c). The case (b) is for R$_{0.25}$E$_{0.75}$ where only the site $a$ is occupied by the R atom (Ru or Rh) the others sites being empty  space or Ag atoms. The case (c) is for R$_{0.50}$E$_{0.50}$. The R atoms are located at sites $a$ and $b$ whereas $c$ and $d$ are empty sites or Ag atoms.}
\end{figure}

We have used the {\it ab initio} method called Plane Waves Self Consistent Field \cite{PWscf}. This method, based on density functional theory \cite{DFT}, is restricted here to use the generalized gradient approximation of Perdew-Burke-Ernzerhof \cite{GGA-PBE} (GGA-PBE) functional.

The studied systems are modelled using slab geometry. Our model consists of various geometrical configurations divided in supported (noted by O) (Fig. 1 a.1), alloyed (Fig. 1 a.2) and buried (Fig. 1 a.3) thin films on Ag(001) substrate. Supported atoms are modelled by R$_{x}$E$_{1-x}$, where $x=0.25$ and 0.50 and R(E) represents the Ru or Rh atoms (empty space). Alloyed atoms are modelled by R$_{x}$Ag$_{1-x}$, where $x$ is the same as above. The unit cell has 4 inequivalent atoms per plane. These geometrical configurations are showed in Fig. $1(b)$ and $1(c)$. Also we studied 1 and 2 monolayers (ML). The geometrical configurations consist of 7 metallic layers, where 5 layers  correspond to Ag(001) substrate and 1 adlayer on each side of the slab. These adlayers represent the supported, alloyed atoms and full monolayers depending of the case. The metallic layers are separated by 7 layers of empty space, which is sufficent to prevent the interaction between the slabs and to vanish the charge at the central layer of empty space. All atoms are located at the ideal positions of a (001) slab with lattice parameter of Ag(001) (4.09$\AA$), however we determined the atomic position  using force minimization. The calculation are performed using the Monkhorst-pack scheme to define the $k$ points for each slab with a grid of $16\times16\times1$. A cut-off energy of 36 Ryd (489eV) was used for the plane waves expansion of the pseudowave function (560Ryd for the charge density and potential). The interlayer distances are relaxed until the absolute force is less than 0.001 Ryd/a.u.

\section{Results}

The aim of this paper is determinate the electronic and structural properties of Ru and Rh thin films on Ag(001) using an {\it ab initio} method, the studied systems are presented in Fig. \ref{fig:Geometry}, and 1(2) full monolayer(s), in order to explain the absence of magnetism in experimental results of Honolka\cite{Honolka} {\it et al.}.

\subsection{Paramagnetic state}

\begin{figure}[!]
\vspace{0.30cm}
\includegraphics[scale=0.33]{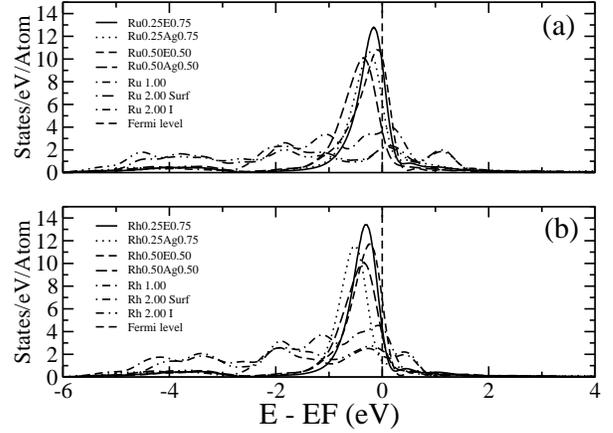}
\caption{\label{fig:NMLDOS}
Paramagnetic local density of states (LDOS) for Ru (a) and Rh (b) atoms (only $d$ component) in Ru and Rh thin films on Ag(001) substrate. 
We present the LDOS for Ru and Rh atoms for each system in alloyed and supported systems with $x=0.25$, $x=0.50$, 
1 and 2 ML (surface (Surf.) and Interface(I)).
 Fermi level is dotted line.}
\end{figure}
\begin{table}[b!]
\caption{ \label{tab:Para}
Values of {\it S} factor for each system. The value for atoms supported (O), alloyed (A), buried (B) surface (Surf) and interface (Inter). 
The letters a b and c are the atomic positions.
}
\begin{ruledtabular}
\begin{tabular}{lrr} 
                &       Ru &          Rh S/A/B \\
\hline
0.25 a O/A      &       2.91/1.96 &        1.46/0.47 \\
0.50 a O/A/B     &    3.05/2.18/1.28 &    2.23/1.27/0.54 \\
1.00 O/B &     1.08/1.08 &         1.43/1.12 \\
2.00 Sur(Inter) &       0.64(0.67) &        0.48(0.55) \\
\end{tabular}
\end{ruledtabular}
\end{table}

\begin{table}
\caption{\label{tab:Energies} Difference of energy (in meV/Atom) between magnetic and non magnetic state for
 Ru and Rh thin films on Ag(001), in the magnetic and non magnetic state. 
The capital letter represent Supported (O), Alloyed (A) and Buried (B).}
\begin{ruledtabular}
\begin{tabular}{lrr}
                      &     Ru    &                   Rh \\
\hline
0.25 O/A   &       -29.86/-12.71/ &       -49.97/-25.82/ \\
0.50 O/A/B & -69.12/-32.20/-11.15 & -121.05/61.23/-13.85 \\
1.00 O/B   &        -33.67/-20.71 &        -86.06/-61.24 \\
2.00 O    &                  0.00 &               105.72 \\
\end{tabular}
\end{ruledtabular}
\caption{\label{tab:MagMom} Magnetic moments (in $\mu_{B}$) for Ru and Rh atoms in 
Rh and Ru thin films on Ag(001) substrate.The capital letter represent supported (O), alloyed (A) and
buried (B).}
\begin{ruledtabular}
\begin{tabular}{lrr}
          &        Ru &             Rh \\
\hline
0.25 O/A/S &     2.49/1.86/ & 2.00/1.80/0.85 \\
0.50 O/A/B & 2.56/2.12/1.48 & 1.99/1.79/1.15 \\ 
1.00 O/B   &      2.07/1.65 &      1.57/1.40 \\ 
2.00 O     &      0.01/0.00 &      1.75/1.51 \\ 
\end{tabular}
\end{ruledtabular}
\caption{\label{tab:Distances} Distances (in $\AA$) between R atoms and Ag(001) substrate,
in the magnetic ($z_{M}$) and non magnetic ($z_{NM}$) case.The O and A represent the Supported and 
Alloyed, respectively.}
\begin{ruledtabular}
\begin{tabular}{lrrrr}
        &\multicolumn{2}{l}{Ru} & \multicolumn{2}{l}{Rh}\\
        &                O &  A &          O & A        \\
        & $z_{NM}$/$z_{M}$ &    $z_{NM}$/$z_{M}$ \\
\hline
0.25 & 1.60/1.72 & 1.91/1.94 & 1.62/1.85 &  1.91/2.03 \\    
0.50 & 1.73/1.80 & 1.90/1.96 & 1.74/1.91 & 1.91/2.05 \\     
1.00 & 1.98/1.99 &          &           1.67/1.74 & \\ 
\end{tabular}
\end{ruledtabular}
\end{table}

The Stoner model\cite{Mohn} is able to reveal the magnetic order in $3d$ transition metals and their alloys, with few values of the system to consider. The needed values to indicate the magnetic order are the paramagnetic Local Density of States (LDOS) and the Stoner parameter ($I_{S}$). Therefore, the Stoner model is able to describe ferromagnetism, weak ferromagnetism, superparamagnetism and paramagnetism. 

We report the paramagnetic LDOS (only $d$ component) for Ru and Rh atoms, see Fig. \ref{fig:NMLDOS}.

Following the Stoner model we have the next equation:

\begin{displaymath}
S=I_{S}N_{NM}(E_{F})
\end{displaymath}
where $I_{S}$ is the Stoner parameter and $N_{NM}(E_{F})$ is the value of paramagnetic LDOS at the Fermi level. The $I_{S}$ values are 0.2992 and 0.3264 (in eV) for Ru and Rh, respectively\cite{Janak}. We report the obtained {\it S} values in Table \ref{tab:Para}.

For low coverage, the LDOS presents a narrow and big peak below the Fermi level, this peak decreases when the coverage increases. The LDOS presents a big value at Fermi level which satisfies the Stoner criterion for low coverage, and when the coverage increases the Stoner criterion is not satisfied. The {\it S} factor decreases when the coverage increases.

We can see in Table \ref{tab:Para} that the {\it S} factor (and consequently the magnetic order) changes as a function of the coverage and the type of system (O, A, or B). It means that the system has a magnetic phase transition from strong ferromagnetism  ($S>>1$) to weak ferromagnetic ($2>S>1$) to superparamagnetic ($S \approx 1$) and finally paramagnetic ($S<1$).
 
We can see in Table \ref{tab:Para} that the {\it S} factor for Rh is always less than 2, except for the case of supported surface ordered alloy, where it is slightly bigger than 2. It means that for the case of Rh thin films on Ag(001) present ferromagnetism only for the supported surface ordered alloy. In other cases the system presents non strong ferromagnetism.

In the case of Ru we can numerate as following: the {\it S} factor is bigger than 3 for $x=0.50$ for Supported atoms; {\it S}=2.91, 2.18 and 1.96 for $x=0.25$ Supported, $x=0.50$ Alloyed and $x=0.25$ Alloyed atoms, respectively; {\it S}=1.28, 1.08 and 1.08 for $x=0.25$ Buried, $x=1.00$ Supported and Buried ML, respectively. Ru systems have a phase transition with differences respect to Rh systems.

In both cases (Rh and Ru) for 1 ML, the {\it S} factor is slightly bigger than 1, it means that 1 ML satisfies the Stoner criterion for superparamagnetism or weak ferromagnetism, and 2ML do not satisfy the Stoner criterion ($S<1$).

We can conclude that alloying and buring (interdiffusion) reduce the {\it S} factor and consequently change the magnetic order. We may find ferromagnetism for very low coverage and for supported atoms, and when the coverage increases the magnetic order changes.

\subsection{Magnetic state}

The possible reasons for vanishing magnetism are alloying, relaxation and many body effects, in this section we report the propierties of the magnetic state, in all cases the magnetic state is the ground state. First we define the difference of magnetic energy as the difference between the magnetic and paramagnetic state, it is given by
\begin{displaymath}
\Delta E_{Mag}= E_{M}-E_{NM}
\end{displaymath}
where $E_{M}$ and $E_{NM}$ are the energies in the magnetic and non-magnetic state respectively. The values of $\Delta E_{Mag}$ are presented in Table \ref{tab:Energies}. We report in Table \ref{tab:Distances} the distance (in $\AA$) between R atom and the Ag(001) substrate in the magnetic ($z_{M}$) and non magnetic ($z_{NM}$) case. Also we present the magnetic moment for Ru and Rh atoms in Table \ref{tab:MagMom} for each system.

In almost all cases the magnetic energies are bigger than the thermal energy ($25$ $meV$). The alloying reduces the magnetic moment, this is due to the neighborhood of Ag atoms. The interdiffusion  or buring reduces the magnetic moments. The magnetism increases the distance to the Ag surface, in agreement with magneto-bulk coupling.

The Stoner criterion is not satisfied when $S<1$, however we find magnetic moments. One possible explanation is the enhanced susceptibility that polarize the Ru and Rh atoms.

We cannot comparate our results with previous theoretical works because the distances used here are obtained by fully relaxation or force minimization.

\subsection{Cluster formation}

 Experimental works have presented the cluster formation\cite{Beckmann} and interdiffusion\cite{Beckmann,Chang}. In order to take into account these phenomena, we cosider the energy for 2D surface aggregate and 2D buried aggregate by the following equation:

\begin{equation}
E_{2D} = \frac{1}{2} \big( E_{R_{1.00}} + E_{Ag_{1.00}} \big)
\label{eq1}
\end{equation}
where $ E_{R_{1.00}}$ and $ E_{Ag_{1.00}}$ are the energies for 1ML of R and Ag on Ag(001). Using Eq. (\ref{eq1}) we determine the differences of total energy for each system, and reported in Table \ref{tab:energies}. In both cases Ru and Rh the system more stable is the magnetic buried aggregate, and in both cases the {\it S} factor is almost 1, it means that this type of aggregate satisfies the Stoner criterion for superparamagnetism or weak ferromagnetism, in other words the Stoner criterion is not satisfied for strong ferromagnetism ($S>>1$).

\begin{table}[!]
\caption{\label{tab:energies}
Energy differences (in meV/atom) respect to the ground state (noted by 0.00) in the magnetic
and non magnetic state.}
\begin{ruledtabular}
\begin{tabular}{lrr}
System  &        NM & Mag \\
\hline
Ru      &           &     \\
Surface Ordered Alloy & 230.20 & 189.10 \\
Surface 2D Aggregate  &  93.89 &  82.74 \\
Buried Ordered Alloy  & 114.69 &  93.20 \\
Buried 2D Aggregate   &  10.35 &   0.00 \\
\hline
Rh      &           &     \\
Surface Ordered Alloy & 114.74 &  65.76 \\
Surface 2D Aggregate  & 117.73 &  66.53 \\
Buried Ordered Alloy  &  37.91 &  24.70 \\
Buried 2D Aggregate   &  30.63 &   0.00 \\
\end{tabular}
\end{ruledtabular}
\end{table}

\section{Conclusion}

Following the experimental work of Honolka\cite{Honolka} about Ru and Rh overlayers films on Ag(001) 
substrate\cite{Honolka}, we have performed a theoretical study of the Ru and Rh 
thin films on Ag(001), by means of an {\it ab initio} method. In order to 
explain the experimental results, i.e. the absence of magnetism, we considered 
two out of the three mean possible causes: relaxation and the alloying effects. 
However, magnetism occurred. It seems then that the origin of the discrepancy 
between the mentioned experiment and current theory must be sought in other causes, 
for example, many-body effects, spin fluctuations, steps among others. 
The alloying reduces the magnetic energy by a factor of 2, and the  magnetic 
moments per atom as well. The magnetism increases the distance between the atoms, 
in agreement with the magneto-bulk coupling. We have used the Stoner model to reveal 
the magnetic order and we have found that the magnetic order is a coverage function, 
and that the systems have a magnetic phase transition in coverage function. 
For very low coverage the system is ferromagnetic, and when the coverage 
increases the magnetic order goes to weak ferromagnetic, superparamagnetic and finally paramagnetic, 
successively. By considering the surface aggregate formation and buried aggregate, we 
determine the difference of total energy and the ground state, in both cases is the 
magnetic 2D buried aggregate, and this aggregate does not satisfy the Stoner criterion. 
The two main conclusion are: (1) the Ru and Rh thin films have a magnetic phase 
transition as a function of the coverage and (2) the 2D buried aggregate is a  
more stable system.



\begin{thebibliography}{10}

\bibitem{Honolka}{J. Honolka, K. Kuhnke, L. Vitali, A. Enders, K. Kern, S. Gardonio, C. Carbone,
S. R. Krishnakumar, P. Bencok, S. Stepanow and P. Garbardella, Phys. Rev. B {\bf 76}, 144412 (2007).}

\bibitem{Schmitz}{P. J. Schmitz, W. Y. Leung, G. W. Graham and P. A. Thiel, Phys. Rev. B {\bf 40}, 11477 (1989).}

\bibitem{Li}{H. Li, S. C. Wu, D. Tian, Y. S. Li, J. Quinn and F. Jona, Phys. Rev. B {\bf 44}, 1438 (1991).}

\bibitem{Mulhollan}{G. A. Mulhollan, R. L. Fink and J. L. Erskine, Phys. Rev. B {\bf 44}, 2393 (1991).}

\bibitem{Liu}{C. Liu and S. Bader, Phys. Rev. B {\bf 44}, 12062 (1991).}

\bibitem{Cox1}{A. J. Cox, J. G. Louderback and L.A. Bloomfield, Phys. Rev. B {\bf 71}, 923 (1993).}

\bibitem{Cox2}{A. J. Cox, J. G. Louderback, S. E. Apsel and L. A. Bloomfield, 
Phys. Rev. B {\bf 49}, 12295 (1994). }

\bibitem{Chang}{S. L. Chang, J. M. Wen, P. A. Thiel, S. G\"unther, J. A. Meyer and J. Behm, 
   Phys. Rev. B {\bf 53}, 13747 (1996).}

\bibitem{Beckmann}{H. Beckmann and G. Bergmann, Phys. Rev. B {\bf 55}, 14350 (1997).}

\bibitem{Lin}{ Tao Lin, M. A. Tomaz, M. M. Schwickert and G. R. Harp, Phys. Rev. B {\bf 58}, 862 (1998).}

\bibitem{Chado}{I. Chado, F. Scheurer and J. P. Bucher, Phys. Rev. B {\bf 64}, 094410 (2001).}

\bibitem{Zhu}{M. J. Zhu, D. M. Bylander and L. Kleinman, Phys. Rev. B {\bf 43}, 4007 (1991).}

\bibitem{Eriksson}{O. Eriksson, R. C. Albers adn A. M. Boring, Phys. Rev. Lett. {\bf 66}, 1350 (1991).}

\bibitem{Blugel92}{S. Bl\"ugel, Phys. Rev. Lett. {\bf 68}, 851 (1992).}

\bibitem{Wu}{R. Wu and A. J. Freeman Phys. Rev. B {\bf 45}, 7222 (1992).}

\bibitem{Blugel95}{S. Bl\"ugel, Phys. Rev. B {\bf 51}, 2025 (1995).}

\bibitem{Turek}{I. Turek, J. Kudrnovsk\'y, M Sob, V. Drchal and P. Weinberger, Phys. Rev. Lett. {\bf 74}, 2551 (1995).}

\bibitem{Wildberger}{K. Wildberger, V. S. Stepanyuk, P. Lang, R. Zeller and P H. Dederichs, 
Phys. Rev. Lett. {\bf 75}, 509 (1995).}

\bibitem{Stepanyuk1}{V. S. Stepanyuk, W. Hergert, P. Rennert, K. Wildberger, R. Zeller and P. H. Dederichs,
Phys. Rev. B {\bf 59}, 1681 (1999).}

\bibitem{Stepanyuk2}{ V. S. Stepanyuk, W. Hergert, K. Wildberger, R. Zeller, P. H. Dederich, Phys. Rev. B {\bf 53}, 2121 (1996).}

\bibitem{Bazhanov}{D. I. Bazhanov, W. Hergert, V. S. Stepanyuk, A. A. Katsnelson, P. Rennert, K. Kokko and
C. Demangeat, Phys. Rev. B {\bf 62}, 6415 (2000).}

\bibitem{Cabria}{I. Cabria, B. Nonas, R. Zeller and P. H. Dederichs, Phys. Rev. B {\bf 65}, 054414 (2002). }

\bibitem{Bellini}{V. Bellini, N. Papakolaou, R. Zeller and P. H. Dederichs, Phys. Rev. B {\bf 64}, 094403 (2001).}

\bibitem{Wahl}{ P. Wahl, L. Diekh\"oner, M. A. Schneider, L. Vitale, G. Wittich and K Kern, 
Phys. Rev. Lett. {\bf 93}, 176603 (2004).}

\bibitem{Mokrousov}{Y. Mokrousov, G. Bihlmayer, S. Heinze and S. Bl\"ugel, Phys. Rev. Lett. {\bf 96}, 147201 (2006).}

\bibitem{PWscf}{S. Baroni, A. Dal Corso, S. de Girancoli, and P. Gianozzi,
http://www.pwscf.org.}

\bibitem{DFT}{P. Hohenberg and W. Kohn, Phys. Rev. B {\bf 136}, B864 (1964); 
W. Kohn and L. J. Sham, Phys. Rev. B {\bf 140}, A1133 (1965).}

\bibitem{GGA-PBE}{J.P. Perdew, K. Burke, and M. Ernzerhof, Phys. Rev. Lett. {\bf77 },
3865 (1996). }

\bibitem{Janak}{J. F. Janak, Phys. Rev. B {\bf 16}, 255 (1977).}

\bibitem{Stoner}{E. C. Stoner Proc. R. Soc. Lond. A {\bf 169 }, 339 (1939).}

\bibitem{Mohn}{P. Mohn, Magnetism in Solid State, An Introduction, Springer-Verlag (Germany) 2003. }


\end{thebibliography}
\end{document}